\documentclass[showpacs,aps,prl,twocolumn]{revtex4-1}
\usepackage{latexsym}
\usepackage{amsfonts}
\usepackage{amssymb}
\usepackage{amsmath}
\usepackage{graphicx}
\usepackage{color}

\begin{document}
\title{Quantum Hall effect in polycrystalline graphene: The role of grain boundaries}
\author{Aron W. Cummings$^{1}$, Alessandro Cresti$^{2}$,  and Stephan Roche$^{1,3}$}
\affiliation{$^1$ICN2 - Institut Catal\`a de Nanoci\`encia i Nanotecnologia, Campus UAB, Bellaterra, 08193 Barcelona, Spain}
\affiliation{$^2$IMEP-LAHC (UMR CNRS/INPG/UJF 5130), Grenoble INP, Minatec, 3, Parvis Louis N\'eel, CS 50257, F-38016, Grenoble, France}
\affiliation{$^3$ICREA - Instituci\'o Catalana de Recerca i Estudis Avan\c{c}ats, 08010 Barcelona, Spain}
\date{\today}

\begin{abstract}
We use numerical simulations to predict peculiar magnetotransport fingerprints in polycrystalline graphene, driven by the presence of grain boundaries of varying size and orientation. The formation of Landau levels is shown to be restricted by the polycrystalline morphology, requiring the magnetic length to be smaller than the average grain radius. The nature of localization is also found to be unusual, with strongly localized states at the center of Landau levels (including the usually highly robust zero-energy state) and extended electronic states lying between Landau levels. These extended states percolate along the network of grain boundaries, resulting in a finite value for the bulk dissipative conductivity and suppression of the quantized Hall conductance. Such breakdown of the quantum Hall regime provoked by extended structural defects is also illustrated through two-terminal Landauer-B\"uttiker conductance calculations, indicating how a single grain boundary induces cross linking between edge states lying at opposite sides of a ribbon geometry.
\end{abstract}
\maketitle

{\it Introduction.}
Massless Dirac fermions in graphene \cite{CastroNeto2009} exhibit remarkable transport characteristics such as Klein tunneling \cite{Katsnelson2006}, weak antilocalization \cite{McCann2006}, and a half-integer quantum Hall effect (QHE) \cite{Novoselov2005, Zhang2005, Novoselov2006}. Such peculiarities stem from the pseudospin degree of freedom, which is embedded in the wave function symmetry of graphene and brings supplementary quantum interferences through the Berry's phase. In high magnetic fields, this results in an unconventional Landau level (LL) spectrum with energies given by $E_{n} = {\rm sgn}(n) \sqrt{2\hbar{v_{F}}^{2}eB|n|}$, the presence of a zero-energy LL, and a quantized Hall conductivity $\sigma_{xy} = 4e^{2}/h \times (n+1/2)$ \cite{Novoselov2005, Zhang2005, Goerbig2011} that is observable at low magnetic field and room temperature \cite{Novoselov2007}. A variety of theoretical \cite{Sheng2006,Evers2008,Ortmann2013,Gattenlohner2014} and experimental \cite{Bolotin2008,Guillemette2013,Nam2013} studies have also suggested the existence of quantum critical states that make graphene insensitive to disorder-induced localization at the Dirac point, both in the presence and absence of magnetic fields. This makes graphene an intriguing candidate for next-generation electronic applications \cite{Novoselov2012}, and as a material for high-precision quantum resistance standards \cite{Janssen2013}.

Currently, chemical vapor deposition (CVD) is the best approach for the growth of wafer-scale graphene, but this process results in a polycrystalline material, where pristine graphene grains of different orientations are stitched together by a disordered network of interconnected grain boundaries \cite{Huang2011,Kim2011}. These grain boundaries (GBs) can dominate the electrical properties of polycrystalline graphene, making them undesirable for electronics applications \cite{Yazyev2010,Dinh2013,Yu2011,Tsen2012}. The precise characterization of this material has thus become crucial from an application perspective. In this regard, the particularly complex and tunable morphology of polycrystalline graphene presents unanswered questions concerning the formation of LLs and the conditions for observation of the QHE.

In this work, we provide theoretical insight into intrinsic magnetotransport phenomena in polycrystalline graphene. Using efficient order-$N$ methods, we compute the density of states (DOS) and dissipative conductivity of polycrystalline graphene samples under a perpendicular magnetic field. The finite size of the grains is shown to strongly suppress the formation of LLs, where the zero-energy LL emerges only when the dimensionless parameter $\kappa = r_{G}/\ell_{B} \geq 1$, with $r_{G}$ the average grain radius and $\ell_B = \sqrt{\hbar/eB}$ the magnetic length. Additionally, in contrast to pristine graphene, the nature of the transport is completely reversed by the polycrystallinity. States at the center of the LL, usually robust against localization, are blocked from propagating across GBs and remain confined inside the grains. Moreover, states between LLs form a bulk percolating network along the GBs. As a result, the dissipative conductivity remains finite and the Hall conductivity quantization is correspondingly suppressed. This picture is supported by two-terminal conductance simulations of a wide graphene ribbon with a transverse line defect, which mimics a GB. These features contrast with the usual characteristics of the metal-insulator transition in exfoliated graphene, and could be experimentally tuned by controlling the grain size distribution during CVD growth of polycrystalline samples.

\begin{figure}[htbp]
\resizebox{8cm}{!}{\includegraphics{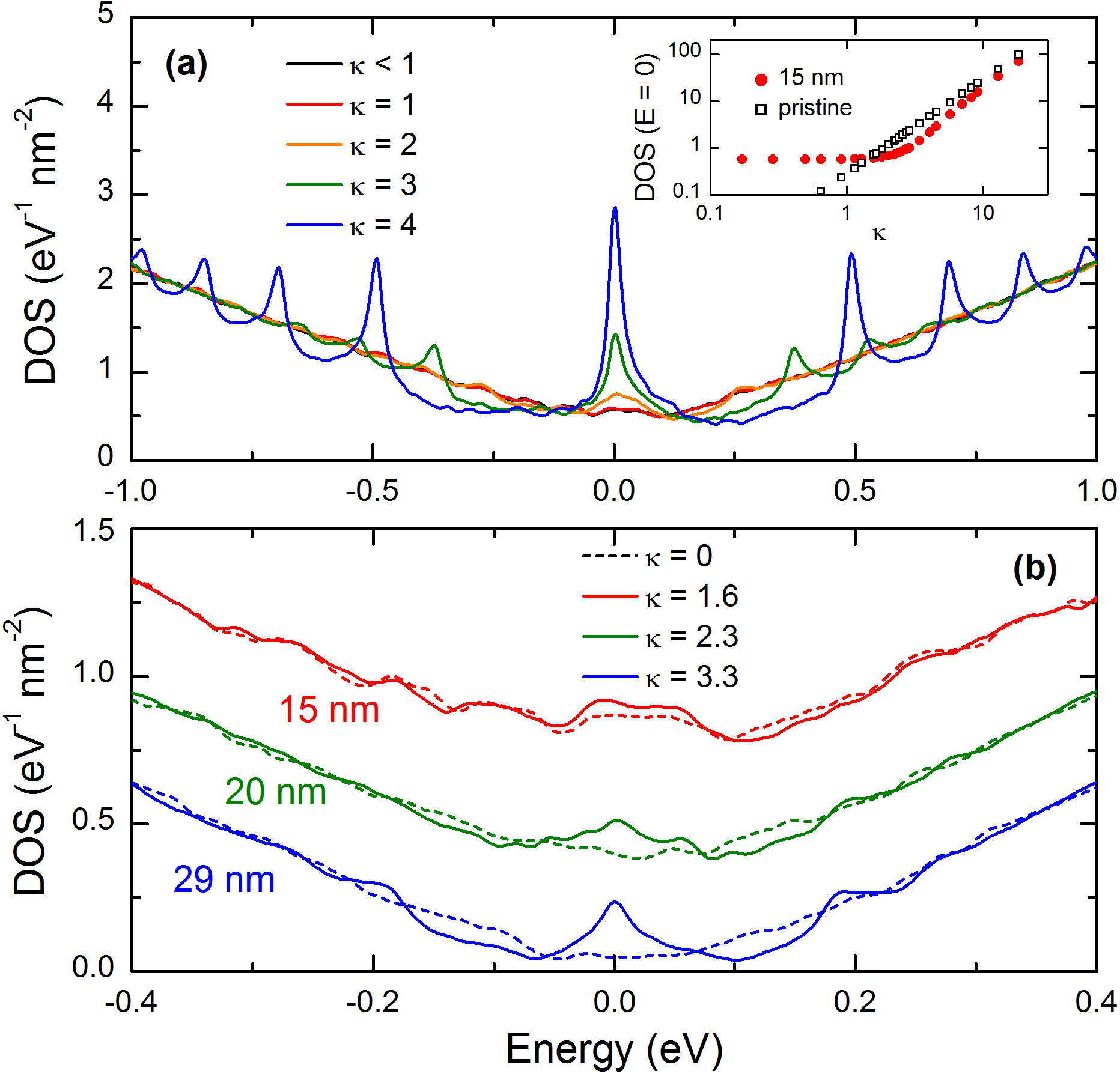}}
\caption{(Color online) (a) Main frame: DOS of a polycrystalline sample with 15 nm average grain diameter and magnetic field strengths given by $\kappa \in [0,4]$. Inset: DOS at the Dirac point for the polycrystalline sample with varying $\kappa$ (solid circles) and for pristine graphene at corresponding magnetic fields (open squares). A broadening of 13 meV was used in both cases. (b) DOS of polycrystalline samples of different grain diameters. Solid (dashed) lines are for a fixed magnetic field of 32 T (0 T). Curves are offset for clarity.}
\label{Fig1}
\end{figure}

{\it System description and simulation methodology.}
The starting point of our study is the simulation of electronic transport in large-area polycrystalline graphene samples, containing millions of atoms and consisting of varying grain misorientation angles, realistic carbon ring statistics, and unrestricted GB structures, based on the method reported in Ref. \cite{Kotakoski2012}. The samples are described with a tight-binding Hamiltonian $\hat{H}$, with the external magnetic field modeled as a standard Peierls phase factor on the hopping elements of $\hat{H}$ \cite{Luttinger1951}. The Kubo conductivity $\sigma_{xx}$ is calculated with an order-$N$, real space approach \cite{Roche1}, which is well suited for large disordered low-dimensional systems. The scaling properties of $\sigma_{xx}$ are followed numerically through wave packet dynamics as
\begin{equation}
\sigma_{xx}(E)=e^{2} \rho(E) \lim_{t \rightarrow \infty} \frac{d}{dt}\Delta X^{2}(E,t),
\end{equation}
where $\rho(E)$ is the DOS, $\Delta X^{2}(E,t) = Tr[\delta(E-\hat{H})|\hat{X}(t)-\hat{X}(0)|^2]/Tr[\delta(E-\hat{H})]$ is the mean square displacement of the wave packet, and $\hat{X}(t)$ is the position operator in the Heisenberg representation. Short (long) time evolution is calculated with a time step of 0.25 fs (0.5 ps). Calculations are performed on systems containing over ten million carbon atoms, corresponding to sizes larger than $500 \times 500~{\rm nm}^{2}$.

For the two-terminal conductance we consider a graphene ribbon with a transverse line defect consisting of pentagonal and octagonal carbon rings, as shown in the inset of Fig. \ref{Fig4}(a). This type of defect has been widely investigated \cite{GUN_PRL106, BAH_PRB83, SON_PRB86, KOU_ACSN5, OKA_JPSJ80, GUN_JVSTB30, LIN_PRB84, IHN_PRB88, YAO_PRB88, ALEX_NL2014} and represents a prototype of the line defects found at the GBs. To demonstrate the generality of our results, we also consider a tilt GB consisting of pentagonal and heptagonal carbon rings, see the inset of Fig. \ref{Fig4}(b). To simulate electron transport we use the Green's function formalism \cite{CRE_EPJB53}, which gives the two-terminal conductance in terms of the Landauer-B\"uttiker formula
\begin{equation}
	G(E) = (2e^2/h) {\rm Tr} [G^R(E)\Gamma^{\rm (S)}(E)G^A(E)\Gamma^{\rm (D)}(E)],
\end{equation}
where $\Gamma^{\rm (S/D)}$ are the rate operators for the source and drain contacts and $G^{R/A}$ are the retarded and advanced Green's functions. 
From the lesser Green's function $G^<$ we extract the local density of occupied states on the $i$th atom as
\begin{equation}
	\rho_{i} (E) = \Im{\rm m} [G^<(E)]_{ii}/(2\pi).
\end{equation} 
This quantity is a valuable tool to analyze how electrons injected from the source distribute within the system. We note that our tight-binding models do not include effects such as spin-orbit coupling or particle-particle interactions, which may have a quantitative impact on the results presented here. However, we believe that the qualitative nature of our results and the general physical mechanisms at play will remain intact with the inclusion of these effects.

{\it Results and discussion.} We first investigate the effect of polycrystallinity on the formation of LLs in samples with different average grain diameters \cite{SM}. Figure \ref{Fig1}(a) shows the evolution of the DOS of a sample with an average grain diameter of 15 nm as the magnetic field varies from 0 to $\kappa = 4$. The formation of LLs is clearly suppressed for low magnetic fields, while for $\kappa = 4$ the DOS begins to resemble that of pristine graphene, with multiple LLs at energies proportional to $\sqrt{B|n|}$. This transition is highlighted in the inset of Fig. \ref{Fig1}(a), which shows the DOS at the Dirac point for the 15-nm sample and for pristine graphene. For $\kappa < 1$, the zero-energy LL cannot form and the zero-energy value of the DOS is dictated by the impurity states induced by the GBs. For $\kappa > 1$, the DOS begins to increase and approaches that of pristine graphene. The condition $\kappa = 1$ occurs when $\ell_{B} = r_{G}$, marking the point where a classical cyclotron orbit can fit inside an individual grain. In Fig. \ref{Fig1}(b), we plot the DOS of three polycrystalline samples of different grain diameters at a fixed magnetic field of 32 T. Due to their differing grain sizes, $\kappa$ is different for each sample. A larger grain size clearly facilitates a stronger formation of LLs since the condition $\kappa > 1$ is met for smaller magnetic fields. Extrapolating to a sample with an average grain diameter of 500 nm, a magnetic field as small as 170 mT should be enough to obtain a well identified LL-spectrum similar to the one seen in Fig. \ref{Fig1}(a) for $\kappa = 4$ (ignoring the effect of disorder within the grains).

\begin{figure}[htbp]
\resizebox{8cm}{!}{\includegraphics{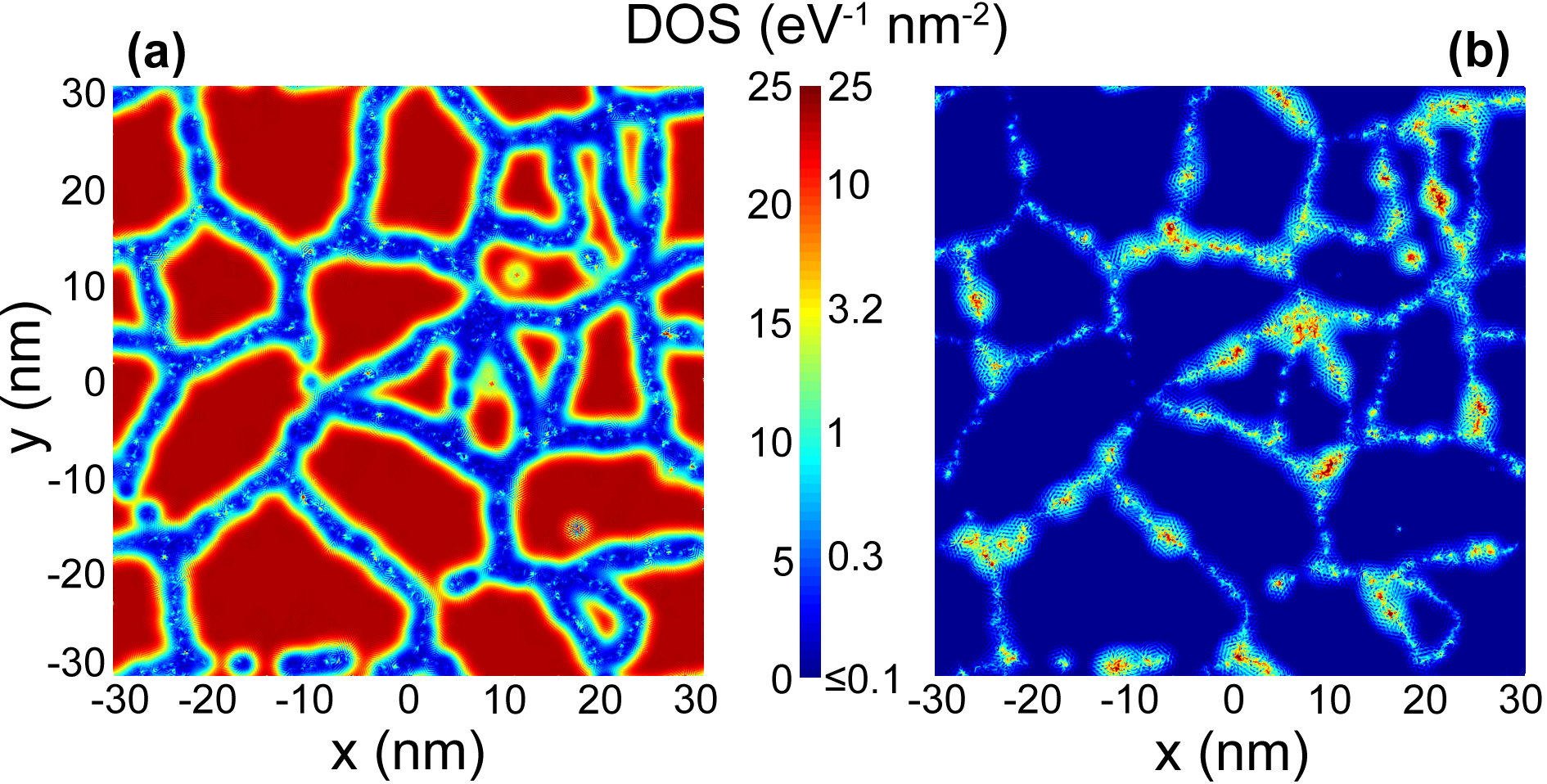}}
\caption{(Color online) (a) Local DOS at the Dirac point for a polycrystalline sample with an average grain diameter of 15 nm. (b) Local DOS for an energy of 0.5 eV located between the zero-energy LL and the first LL. In both panels, $\kappa = 9$.}
\label{Fig2}
\end{figure}

Next we investigate the effect of polycrystallinity in the strong quantum Hall regime, when $\kappa \gg 1$. A peculiar spatial distribution of the electronic states is obtained in this regime, as shown in Fig. \ref{Fig2}, where we plot (for $\kappa = 9$) the local DOS of the 15-nm sample (a) at the Dirac point, and (b) at an energy halfway between the zero-energy LL and the first LL. At the Dirac point, states are mainly confined inside the grains, with the GBs acting as strong impeding barriers, disconnecting electronic propagation from grain to grain. Meanwhile, between LLs the DOS remains finite with electronic states residing on and around the GBs, forming an extended network throughout the bulk of the sample. This particular spatial distribution of states also applies to higher energies at and between LLs, and plays an important role in the transport properties of the polycrystalline sample.

\begin{figure}[htbp]
\resizebox{8cm}{!}{\includegraphics{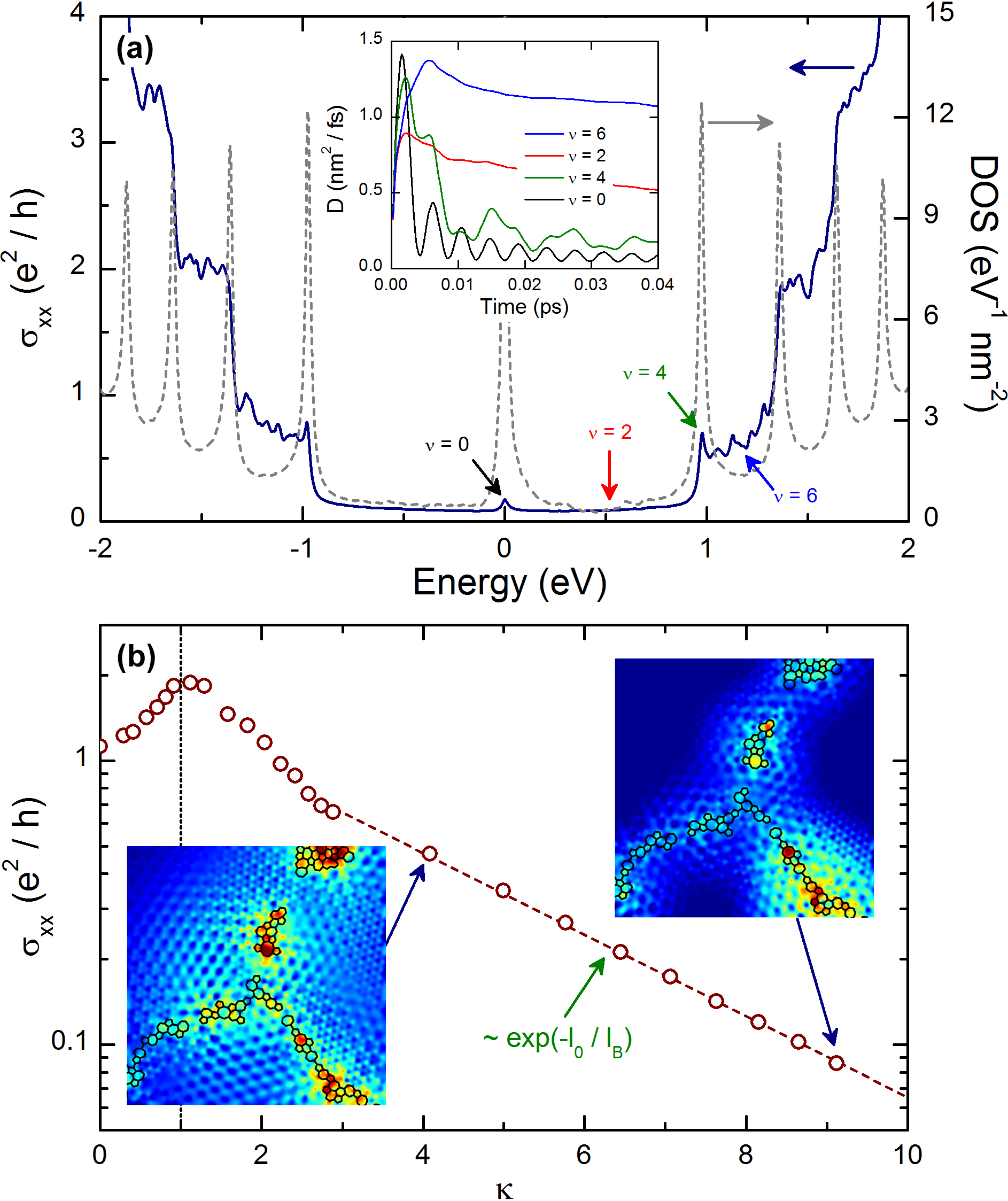}}
\caption{(Color online) (a) Kubo conductivity (left axis) and superimposed DOS (right axis) of a polycrystalline sample with 15 nm average grain diameter and $\kappa = 9$. The conductivity has been calculated at a simulation time of 10 ps. Inset: Time-dependent diffusion coefficient at selected energies ($\nu = 0,2,4,6$) marked by arrows in the main frame. (b) Magnetic field dependence of the conductivity at $\nu = 2$, indicating a transition to a percolating regime at $\kappa > 1$. The insets show a zoom-in of the LDOS at $\nu = 2$ for $\kappa = 4$ and $\kappa = 9$, where the black lines show the location of the GBs.}
\label{Fig3}
\end{figure}

\begin{figure}[t]
\resizebox{8cm}{!}{\includegraphics{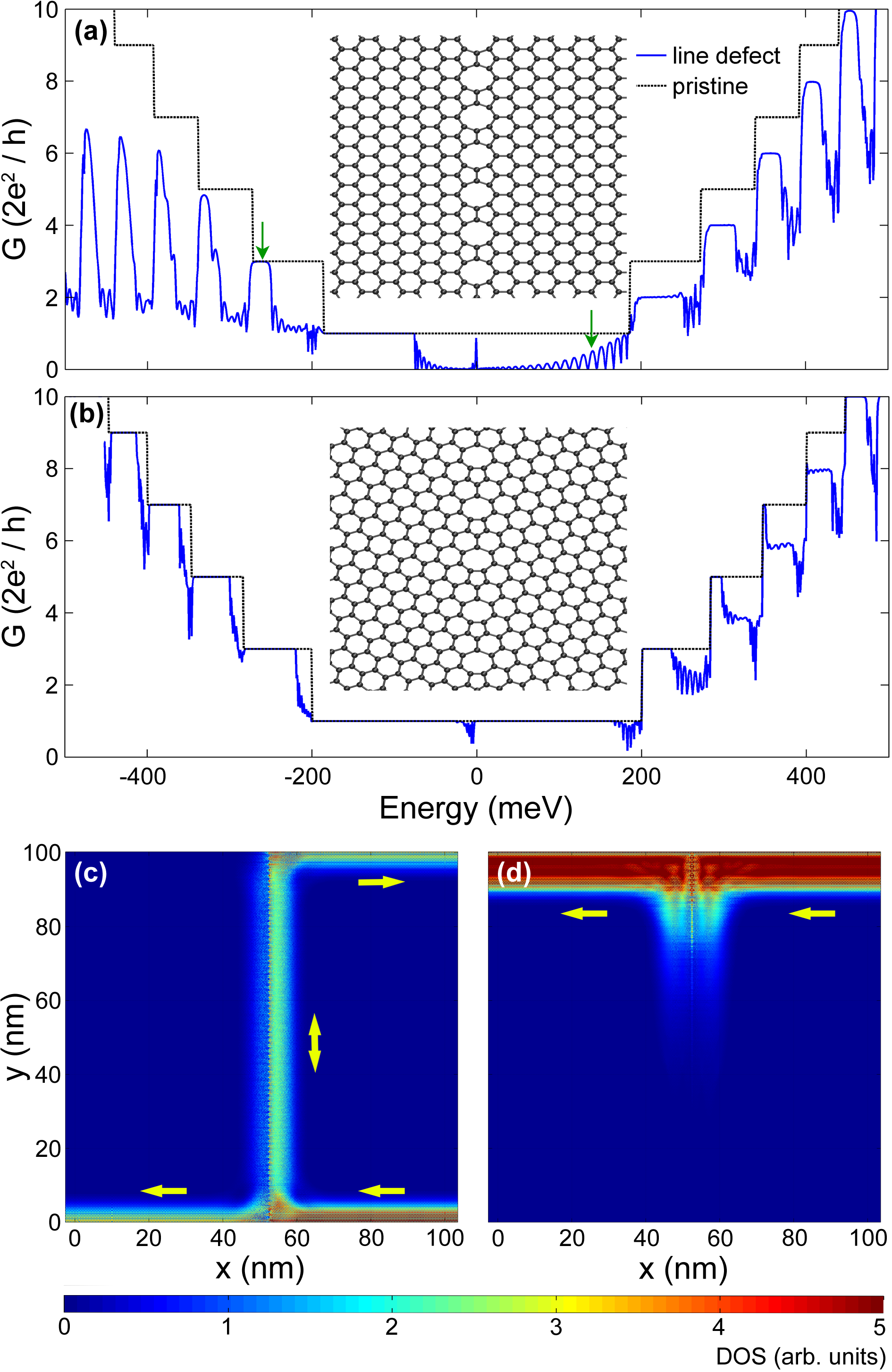}}
\caption{(Color online) (a) Transmission coefficient vs. energy $E$ of a 100 nm wide armchair ribbon with a transverse GB consisting of pentagonal and octagonal rings, shown in the inset, at $B = 40$ T. (b) Same as (a) but with a transverse GB consisting of pentagonal and heptagonal rings. (c) Spatial distribution of the right-injected local DOS at $E = 140$ meV. (d) Same as (c) at $E = -262$ meV.}
\label{Fig4}
\end{figure}

This can be seen in Fig. \ref{Fig3}(a), where we plot the longitudinal conductivity $\sigma_{xx}$ (solid line) for this sample, superimposed with the total DOS (dashed line). The curve for $\sigma_{xx}$ has been taken at a simulation time of $t = 10~{\rm ps}$, allowing the wave packet to explore a large number of grains. In contrast to other forms of disordered graphene \cite{Sheng2006}, the energy dependence of $\sigma_{xx}$ does not reflect that of the LL spectrum. In particular, the conductivity is suppressed at the LLs, while it remains finite between LLs. This situation is opposite that of the conventional QHE, for which states at the center of LLs are robust against localization while bulk states beyond the mobility edges all become localized, enabling both a quantized Hall conductivity and a longitudinal conductivity that qualitatively resembles the DOS.

The nature of the states at and between LLs can be partially revealed by their time-dependent behavior. This is shown in the inset of Fig. \ref{Fig3}(a), where we plot the diffusion coefficient $D(E,t) = \frac{d}{dt}\Delta X^{2}(E,t)$ for energies at the center of two LLs (marked by arrows at $\nu = 0$ and $4$) and energies between LLs (at $\nu = 2$ and $6$). The localized nature of the states at the center of the LLs is clear from the fast decay of $D$. The oscillations are related to the corresponding cyclotron resonances $\omega_{c} = v_{F}/\ell_{B} \times (\sqrt{2(n+1)}-\sqrt{2n})$ \cite{Goerbig2011}. Meanwhile, between LLs, $D$ exhibits a weak time-dependent decay, which is typical for extended states in the weak localization regime \cite{Roche1}. This behavior connects to the finite value of $\sigma_{xx}$ between LLs, where the current is conveyed by states which propagate through the GB network as pictured in Fig. \ref{Fig2}(b). It should be noted that the precise morphology of our samples can have a quantitative effect on our results. For example, larger and better-connected grains result in reduced localization at the LLs, giving a profile of $\sigma_{xx}$ that more closely resembles the DOS \cite{SM}. However, the qualitative nature of our results, and the fundamental role played by the GBs, remains unchanged.

To more clearly understand the transport mechanism associated with the finite bulk conductivity between LLs, in Fig. \ref{Fig3}(b) we plot $\sigma_{xx}$ as a function of $\kappa$, where $\sigma_{xx}$ is chosen at an energy halfway between the zeroth and first LL. For low magnetic fields ($\kappa < 1$) the system exhibits a negative magnetoresistance behavior, highlighting a weak localization effect induced by the GBs. Meanwhile, for $\kappa > 1$, LLs are able to form and the electronic states concentrate along the GB network, as illustrated in Fig. \ref{Fig2}(b). The system then transitions to a strong regime of positive magnetoresistance. In particular, the scaling of the conductivity follows $\sigma_{xx} \propto \exp(-\alpha\kappa) \propto \exp(-l_0/l_B)$, as indicated by the dashed line in Fig. \ref{Fig3}(b). This scaling with magnetic field is indicative of percolation transport through the bulk network of disordered GB states, mediated by hopping between adjacent clusters of non-hexagonal defects. In this transport picture the conductivity scales as $\exp(-R/\xi)$, where $R$ is a characteristic distance between defect states and $\xi$ is the size of the wavefunction associated with each defect state \cite{Ambegaokar1971}. In a strong magnetic field the size of the wavefunction becomes proportional to $l_B$, giving a conductivity that scales as $\exp(-l_0/l_B)$ \cite{Shklovskii1983}. This behavior is visualized in the insets of Fig. \ref{Fig3}(b), where we plot a zoom-in of the LDOS at $\nu = 2$ for $\kappa = 4$ (left inset) and $\kappa = 9$ (right inset). The superposed black lines indicate the position of the GBs. These images illustrate that the states around the GB defects become more localized with stronger magnetic fields, which in turn reduces the hopping between defect sites and thus the conductivity through the GB network. The fit to the data in Fig. \ref{Fig3}(b) gives a value of $l_0 = 2.5$ nm for the 15-nm sample.

To further clarify the effect of individual GBs on transport in the quantum Hall regime, we analyze the two-terminal conductance of a graphene ribbon with a width of 100 nm under a perpendicular magnetic field of $B=40$ T. As shown in Figs. \ref{Fig4}(a) and \ref{Fig4}(b), the pristine ribbon exhibits the usual quantization of the half-integer QHE. The plateaus correspond to energies between LLs, where the current is carried through chiral edge states. In resonance with our results for two-dimensional (2D) graphene, the conductance quantization is partially lost when a transverse line defect is incorporated within the ribbon. This effect, apart from the quantitative details of the conductance, is present in both GBs and is expected to be general and independent of the specific geometry of the GB.

Therefore, we focus on the geometry of Fig. \ref{Fig4}(a) and examine the spectral distribution of injected charge carriers at two representative energies. First we consider $E = 140$ meV, corresponding to a high scattering region. The local DOS of electrons injected from the right side of the system is shown in Fig. \ref{Fig4}(c), where electrons initially flow along the bottom edge, corresponding to the left-moving chiral edge state. When they reach the GB, they are partly transmitted to the left contact and partly deflected along the GB and finally backscattered along the top edge. This cross linking of edge states is at the heart of the loss of the Hall conductance quantization, and has recently been reported in graphene samples decorated with bilayer patches \cite{Chua2014,Schumann2012,Lofwander2013}.

We also consider the energy $E = -262$ meV, which corresponds to a region in Fig. \ref{Fig4}(a), where the conductance remains quantized, i.e., backscattering is absent. The local DOS of injected electrons is shown in Fig. \ref{Fig4}(d). Injected charges flow along the top edge (the chirality of the edge states is opposite for negative energies) and upon reaching the line defect, only partially penetrate into the bulk while continuing along the top edge to the left contact. The poor penetration along the line defect prevents charge from reaching the other edge, thus suppressing backscattering. We note that this suppression of backscattering is absent from our bulk calculations, and thus is likely a feature of the periodicity and band structure of these particular GBs \cite{SM}.

While GBs in CVD graphene have more complex geometries than the line defects considered here, our results capture and visualize the main mechanism of the QHE breakdown, complementing our results for 2D graphene. Finally, we would like to remark that the conductance reported in Fig. \ref{Fig4}(a) shows several peculiar features which are specific to the geometry of the line defect considered and whose interpretation is thus deferred to the Supplemental Material \cite{SM} for the interested reader. The richness of the physics seen in our simulations may also find experimental confirmation, since very long and clear line defects have been experimentally reported \cite{LAH_NN5}.

{\it Conclusions.}
While the QHE appears to be robust against strong point scatterers such as hydrogen \cite{Guillemette2013}, the presence of GBs in polycrystalline graphene jeopardizes the emergence of the QHE in two respects. For small magnetic fields, the formation of LLs is limited by the size of the grains and cannot occur until $\ell_{B}<r_{G}$. When LLs are fully developed, the nature of localization remains energy dependent but is strongly altered by the GBs, with bulk transport facilitated for energies between LLs and localization inside the grains at the center of LLs. Together, these effects are expected to suppress the quantized Hall conductance and may help to explain recent quantum Hall measurements of CVD-grown graphene \cite{Shen2011,Nam2013,LaFont2014,Bergvall2014}. This phenomenon has also been illustrated by the cross linking of edge states induced by a single GB in a graphene ribbon geometry. Such peculiar transport features open possibilities for improved structural characterization of CVD-grown polycrystalline graphene, since the average grain size and the density of nonhexagonal defects is directly connected to anomalous transport characteristics in the high magnetic field regime.

The research leading to these results has received funding from the European Union Seventh Framework Programme under Grant Agreement No. 604391 Graphene Flagship. This work was also funded by Spanish Ministry of Economy and Competitiveness under Contract No. MAT2012-33911.

\end{document}